\documentclass[useAMS,usenatbib,usegraphicx]{mn2e}
\usepackage{pstricks}
\usepackage{fixltx2e}

\def\mnras{MNRAS}  
\def\apj{ApJ}      
\def\apjl{ApJL}    
\def\aap{A\&A}     
\def\aj{AJ}        
\def\spose#1{\hbox to 0pt{#1\hss}}
\def\ltsimm{\mathrel{\spose{\lower 3pt\hbox{$\sim$}}
	\raise 2.0pt\hbox{$<$}}}

\title[Stripping of galactic haloes using the $k$-$\epsilon$ model]{Ram pressure stripping of the hot gaseous haloes of galaxies using the $k$-$\epsilon$ sub-grid turbulence model}

\author[J. L. Close et al.]{J. L. Close$^{1}$\thanks{E-mail: py08jlc@leeds.ac.uk}, J. M. Pittard$^{1}$, T. W. Hartquist$^{1}$ and S. A. E. G. Falle$^{2}$\\
$^{1}$School of Physics and Astronomy, University of Leeds, Leeds, LS2 9JT, UK\\
$^{2}$Department of Applied Mathematics, University of Leeds, Leeds, LS2 9JT, UK}

\begin{document}

\date{\today}

\maketitle

\begin{abstract}
We perform three dimensional hydrodynamic simulations of the ram pressure stripping of the hot extended gaseous halo of a massive galaxy using the $k$-$\epsilon$ sub-grid turbulence model at Mach numbers $0.9$, $1.1$ and $1.9$. The $k$-$\epsilon$ model is used to simulate high Reynolds number flows by increasing the transport coefficients in regions of high turbulence.  We find that the initial, instantaneous stripping is the same whether or not the $k$-$\epsilon$ model is implemented and is in agreement with the results of other studies. However the use of the $k$-$\epsilon$  model leads to five times less gas remaining after stripping by a supersonic flow has proceeded for $10\,$Gyr, which is more consistent with what simple analytic calculations indicate. Hence the continual Kelvin-Helmholtz stripping plays a significant role in the ram pressure stripping of the haloes of massive galaxies. To properly account for this, simulations of galaxy clusters will require the use of sub-grid turbulence models.
\end{abstract}

\begin{keywords}
hydrodynamics -- methods: numerical -- galaxies: haloes -- galaxies: evolution -- galaxies: clusters: general.
\end{keywords}


\section{Introduction}

The properties of galaxies that exist in clusters differ from those of galaxies that exist in the field. This is the so called morphology-density relation \citep{1980ApJ...236..351D}.  A higher local galaxy density is correlated with a larger fraction of late-type (elliptical, E/S0) galaxies \citep{2003MNRAS.346..601G}. It is also associated with a number of other galactic properties, including a decreased rate of star formation \citep{2003ApJ...584..210G} and a higher fraction of red galaxies \citep{2004ApJ...615L.101B}. Also, dwarf elliptical galaxies are more strongly clustered than dwarf irregulars \citep*{1987AJ.....94..251B, 1990A&A...228...42B}. These observational results strongly suggest that a cluster environment shapes the evolution of the galaxies within it.

One possible way to explain the morphology-density relation is ram pressure stripping. This was first explored analytically for a disk shaped gas distribution by \citet{1972ApJ...176....1G}. As a galaxy orbits within a cluster it moves though the inter-cluster medium (ICM) which exerts a ram pressure on the gas in the galaxy. Ram pressure stripping can be separated into two distinct processes, instantaneous stripping and Kelvin-Helmholtz stripping. The instantaneous stripping occurs when the ram pressure is higher than the gravitational force per area on a column of gas and occurs on short time-scales (less than $1\,$Gyr in most cases). Kelvin-Helmholtz stripping is due to the shear force created as the ICM flows past the edge of the galaxy. This induces the Kelvin-Helmholtz instability and allows material to be continually stripped from the galaxy and occurs on longer time scales. Ram pressure stripping has led to tails in a number of galaxies in the Virgo cluster \citep{2005AJ....130...65C, 2011AJ....141..164A}. 

A number of studies have been done to try to quantify different aspects of this effect. In general, the effect of ram pressure stripping is simulated in two ways. Firstly one can perform a "wind tunnel" test. Here the galaxy is placed in a constant wind to simulate the effect of the galaxy moving though the ICM. This allows parameters, like the relative speed of the ICM and galaxy, to be precisely controlled. The other method is to allow the galaxy to orbit within a cluster potential. This gives a more realistic representation of ram pressure stripping and includes tidal effects (which may or may not be desirable).

Many studies have been performed to investigate the effect of ram pressure stripping on spiral galaxies. \citet*{1999MNRAS.308..947A} used a smoothed particle hydrodynamics (SPH) code to simulate a three-dimensional spiral galaxy undergoing ram pressure stripping. Although their results match those of the analytic approach of Gunn \& Gott, it cannot account fully for the observations. \citet{2006MNRAS.369..567R} looked specifically at the role of the inclination angle of the disk and its effect on the morphology of the galaxy, using a grid based code. They found that the mass loss of the galaxy is relatively insensitive to the inclination angle for angles $\lse 60^\circ$. They also noted that the tail is not necessarily pointing in the same direction as the motion of the galaxy. \citet{2009A&A...500..693J} also looked at the role of inclination angle but used a SPH code. They find much the same dependence on inclination angle as \citet{2006MNRAS.369..567R} although they saw no Kelvin-Helmholtz stripping. \citet{2009ApJ...694..789T} allowed the gas in the disk of the spiral galaxy to radiatively cool before being hit by the wind to allow areas of low and high density to develop. They found that gas is stripped more rapidly in the case with cooling as areas of lower density allow gas to also be stripped from the inner regions of the galaxy.

A number of studies have instead been focused on dwarf galaxies. \citet{2000ApJ...538..559M} looked at the stripping of the extended hot gas component using a two-dimensional grid based code. They found that the gas is totally stripped in a typical galactic cluster. \citet{2002MNRAS.336..119M} simulated a dwarf galaxy orbiting the Milky Way with an N-body code to probe the tidal effects on the dwarf galaxy and found a significant mass loss over a period of $7\,$Gyr. \citet{2006MNRAS.369.1021M} used a combination of SPH and N-body simulations to study the combined effects of ram pressure and tidal stripping on dwarf galaxies. They pointed out that tidal effects can alter the morphology of the dwarf galaxy and change the effectiveness of ram pressure stripping. In general this increases effectiveness but tidally induced bar formation can funnel gas towards the centre of the galaxy, making it harder to strip.

In addition to these works a few papers contain results of work specifically on the effect of ram pressure stripping on the hot extended component of the gas. \citet{2008MNRAS.383..593M} simulated a massive galaxy with a hot halo of gas undergoing ram pressure stripping. They performed both wind tunnel tests and simulations allowing the galaxy to orbit within a cluster to include tidal effects. They extended the analytic model of \citet{1972ApJ...176....1G} to the case of a spherical gas distribution and found that their formula fits simulations well when the galaxy represents less than 10\% of the cluster mass. They suggested that at this point tidal effects and gravitational shock heating become important, which their model does not take into account. \citet{2009MNRAS.399.2221B} included both a hot halo and a disc in their model and suggested that the presence of the disk can suppress ram pressure stripping. \citet{2013MNRAS.428..804S} used a grid based code to look at the effects of turbulence within the ISM. They found that this increases mass loss and allows the ICM to penetrate further into a galaxy.


\section{Turbulence Model}

The majority of studies on  ram pressure stripping have been performed
with SPH  codes.  While  these can model  the effect  of instantaneous
stripping,   they    are   known   to    have   difficulty   resolving
Kelvin-Helmholtz         instabilities         (KHIs)        correctly
\citep{2007MNRAS.380..963A}.   Typically studies  done with  SPH codes
show little  to no Kelvin-Helmholtz  stripping, whereas it is  seen in
studies done with grid based codes. However, grid based codes are also
currently unable to reproduce  the turbulence actually involved in ram
pressure stripping.  The turbulent behaviour  of a flow depends on its
Reynolds number, $Re = ur/\nu$, where $u$ is the velocity of the flow,
$r$ is the typical length scale of the flow and $\nu$ is the kinematic
viscosity.  In this work we take $r$ as the radius of the galaxy after
instantaneous  stripping.  For  a fully  ionized, non-magnetic  gas of
density $\rho$ and temperature $T$, the kinematic viscosity

\begin{equation}
\nu    =    2.21\times10^{-15}\frac{T^{5/2}   A^{1/2}}{Z^{4}\rho\,{\rm
ln}\Lambda}\,{\rm g\,cm^2\,s^{-1}},
\end{equation}
where $A$  and $Z$ are  the atomic weight  and charge of  the positive
ions    and    ${\rm   ln}\Lambda$    is    the   coulomb    logarithm
\citep{Spitzer:1956}.

We  can  estimate the  Reynolds  number of  the  ICM  using the  above
equations. Taking  $v=580\,{\rm km \,s^{-1}}$  as the flow  speed past
the galaxy, $r=79$\,kpc as the characteristic size of the galaxy after
instantaneous stripping,  $\rho = 1.7\times10^{-28}\,{\rm g\,cm^{-3}}$
and $T=10^{7}$\,K as the ICM density and temperature, $A \approx 1.0$,
$Z  \approx 1.0$ and  ln$\Lambda\approx 30$,  gives $\nu  \approx 1.35
\,{\rm  cm^2\,s^{-1}}$ and  $Re\approx100$ for  the ICM  gas  for pure
hydrodynamics.

However, the ICM contains a magnetic field, which is typically
of  order  $1\,{\mu}$G  (see  e.g.  \citealt{Carilli:2002}).   If  the
thermal  velocity  is  $\approx  5  \times  10^{7}\,{\rm  cm\,s^{-1}}$
(appropriate  for $10^{7}$\,K  plasma), then  the  particle gyroradius
$\approx 2.5 \times 10^{10}\,$cm.  Since this is small compared to the
size of the largest turbulent eddies, the ions are constrained to move
along the field.   For the tangled fields that  one would expect, this
leads to a  considerable reduction in the viscosity,  which means that
we can  safely assume that we  are dealing with  large Reynolds number
turbulence, certainly too large for direct numerical simulation.

The  only tractable  method to  describe such  flows is  a statistical
approach. We therefore  use a subgrid turbulent viscosity  model in an
attempt  to  calculate  the  properties  of  the  turbulence  and  the
resulting increase  in the transport  coefficients. The $k$-$\epsilon$
model is  widely used  and appropriate in  this context.  The earliest
development  efforts  on this  model  were  by \citet{Chou:1945}.  The
closure     coefficients     were     subsequently     adjusted     by
\citet{Launder:1974} to create the  accepted "standard" model, and its
popularity  has   led  to   it  featuring  extensively   in  textbooks
\citep[e.g.,][]{Pope:2000,Davidson:2004,Wilcox:2006}.   It  introduces
two extra fluid variables: $k$, the turbulent energy per unit mass and
$\epsilon$, the  turbulent dissipation rate per  unit mass. Turbulence
is  modelled through  the use  of $k$  and $\epsilon$  to  calculate a
turbulent  viscosity  which increases  the  transport coefficients  in
regions of high  turbulence. In this paper we  use the term "inviscid"
to   refer  to   simulations  performed   without  the   use   of  the
$k$-$\epsilon$ model.

The equations that determine $k$ and $\epsilon$ are largely empirical,
but  have  been used  to  great success  in  engineering  and in  some
astrophysical         problems        ({\citealt{1994MNRAS.269..607F},
\citealt{2009MNRAS.394.1351P},  \citealt*{2010MNRAS.405..821P}).   For
instance, we found  in \citet{2009MNRAS.394.1351P} that $k$-$\epsilon$
simulations  of the turbulent  ablation of  clouds showed  much better
convergence in  resolution tests  than inviscid simulations,  and that
the  converged  solution in  the  $k$-$\epsilon$  models  was in  good
agreement  with the  highest  resolution inviscid  models. Of  course,
other possibilities and variations  also exist for modelling turbulent
flows. For  instance, \citet{2008ApJ...686..927S} use  a model similar
to  the  $k$-$\epsilon$  model  to  capture  the  Rayleigh-Taylor  and
Richtmyer-Meshkov  instabilities  and  buoyancy-driven  turbulence  in
active galactic nuclei,  and \citet{Schmidt:2011} recently developed a
subgrid model for highly compressible astrophysical turbulence.

The  $k$-$\epsilon$   model  used  has  been   shown  to  give
reasonable agreement with  experiment for turbulent underexpanded jets
with Mach numbers in the range $1-2$ and density contrasts $\approx 2$
(\citealt{Fairweather:2006}).   This gives us  some confidence  in the
the  present simulations  since both  problems involve  the  growth of
turbulent shear  layers and the dimensionless parameters  are not very
different: the Mach  numbers are similar, the density  contrast in the
galactic  shear layer  is $\approx  10$ and  the Reynolds  numbers are
large in both cases.


\section{Hydrodynamics}

\subsection{Method of simulation}

In the $k$-$\epsilon$ model the equations \citep{2009MNRAS.394.1351P} describing continuity, momentum, energy, scalar, turbulent energy and turbulent dissipation are, respectively, 

\begin{equation}
	\frac{\partial \rho}{\partial t} + \nabla \cdot (\rho \bmath{u}) = 0,
	\label{eqn:de1}
\end{equation}
\begin{equation}
	\frac{\partial \rho \bmath{u}}{\partial t} + \nabla \cdot (\rho \bmath{u} \bmath{u}) + \nabla P  - \nabla \cdot \bmath{\tau} =  \rho \bmath{g},
	\label{eqn:de2}
\end{equation}
\begin{equation}
	\frac{\partial E}{\partial t} + \nabla \cdot [(E + P)\bmath{u} - \bmath{u} \cdot \bmath{\tau}] - \frac{\gamma}{\gamma - 1} \nabla \cdot (\mu_T \nabla T) = \rho \bmath{u} \cdot \bmath{g},
	\label{eqn:de3}
\end{equation}
\begin{equation}
	\frac{\partial \rho C}{\partial t} + \nabla \cdot (\rho C \bmath{u}) - \nabla \cdot (\mu_T \nabla C) = 0,
	\label{eqn:de4}
\end{equation}
\begin{equation}
	\frac{\partial \rho k}{\partial t} + \nabla \cdot (\rho k \bmath{u}) - \nabla \cdot (\mu_T \nabla k) = P_t - \rho \epsilon,
	\label{eqn:de5}
\end{equation}
\begin{equation}
	\frac{\partial \rho \epsilon}{\partial t} + \nabla \cdot (\rho \epsilon \bmath{u}) - \nabla \cdot (\mu_\epsilon \nabla \epsilon) = \frac{\epsilon}{k}(C_1 P_t - C_2 \rho \epsilon).
	\label{eqn:de6}
\end{equation}
$\rho$ is  the mass density, $\bmath{u}$  is the velocity,  $P$ is the
thermal  pressure and  $E$ is  the total  energy density  (thermal and
kinetic). $C$ represents any advected scalar. In this case we use only
one, to distinguish between the galactic  gas and the ICM. $C_1 = 1.4$
and $C_2 = 1.94$ are  both constants. $\bmath{g}$ is the gravitational
field. We use  a static gravitational field to  simulate the effect of
dark matter, as described in  section \ref{sec:ic}. The effects of gas
self-gravity are neglected as the gravitational field is assumed to be
dominated by the dark matter.

The effect of $k$ and $\epsilon$ on the other fluid variables is characterised by

\begin{equation}
\label{eqn:td}
	\mu_T = \rho C_\mu \frac{k^2}{\epsilon},
\end{equation}
and

\begin{equation}
	\mu_\epsilon = \frac{\mu_T}{1.3},
\end{equation}
where $C_\mu = 0.09$. $P_t$ is the turbulent production term. Using the summation convention,

\begin{equation}
	P_t = \mu_T \left[ \frac{\partial u_i}{\partial x_j} \left( \frac{\partial u_i}{\partial x_j} + \frac{\partial u_j}{\partial x_i}\right) \right] - \frac{2}{3} \nabla \cdot \bmath{u}(\rho k + \mu_T \nabla \cdot \bmath{u}).
\end{equation}
$\bmath{\tau}$ is the turbulent stress tensor and is defined as

\begin{equation}
	\tau_{ij} = \mu_T \left( \frac{\partial u_i}{\partial x_j} + \frac{\partial u_j}{\partial x_i}\right) - \frac{2}{3} \delta_{ij}(\rho k + \mu_T \nabla \cdot \bmath{u}).
\end{equation}
The equations  simplify to  the familiar inviscid  equations for  $k =
0$.  The values  of $C_1$,  $C_2$ and  $C_\mu$ are  determined  by the
requirement  that a  variety of  experimental results  are  matched by
model results \citep{Dash:1983}. So  they are not free parameters. The
maximum turbulent length  scale must also be chosen,  which limits the
size  of the largest  eddies. We  choose the  predicted radius  of the
galaxy after  instantaneous stripping, $120$\,kpc,  as the appropriate
value. Tests indicate  that the simulations are not  very sensitive to
this  parameter,  with  the  biggest  differences  being  in  the  far
downstream  structure of  the wake.  Importantly the  turbulent mixing
layer around the  galaxy is not affected by our  choice of the maximum
turbulent length scale, and therefore the rate of stripping within our
model is robust.

The calculations were performed with the hierarchical adaptive mesh refinement (AMR) code, \texttt{MG}. The code uses the Godunov method, solving a Riemann problem at each cell interface, using piece-wise linear cell interpolation and MPI-parallelization. The equations \ref{eqn:de1}-\ref{eqn:de6} are solved using the second order upwind scheme described in \citet{1991MNRAS.250..581F} for the hyperbolic terms combined with a centred difference for the diffusive terms. A hierarchy of $N$ grids levels, $G_0 \cdots G_{N-1}$, is used, and the mesh spacing for $G_n$ is $\Delta x/2^{n}$, where $\Delta x $ is the cell size for the coarsest level, $G_0$.  $G_0$ and $G_1$ cover the entire domain, but finer grids need not do so.  Refinement is on a cell-by-cell basis and is controlled by error estimates based on the difference between solutions on different grids, i.e. the difference between the solutions on $G_{n-1}$ and $G_n$ determine refinement to $G_{n+1}$. The efficiency is increased by using different time steps on different grids, such that $G_{n}$ undergoes two steps for each step on $G_{n-1}$. In these calculations $G_0$ had 50 cells per side and there were 5 additional levels of refinement, giving a maximum resolution of 1600 cells per side, corresponding to a minimum mesh spacing of $6\,$kpc.

\subsection{Initial conditions}
\label{sec:ic}

\begin{table}
	\caption{Galaxy model parameters}
	\label{table:1}
	\centering
	\begin{tabular}{l c}
		\hline
		Variable & Value \\
		\hline
   		$M_{200}$ & $2 \times 10^{12}$M$_{\sun}$ \\
   		$c_{200}$ & $4$ \\
   		$f_b$ & $0.141$ \\
   		$r_{s}$ & $79$kpc \\
		\hline
	\end{tabular}
\end{table}

\begin{table}
	\caption{ICM parameters}
	\label{table:2}
	\centering
	\begin{tabular}{l c}
		\hline
		Variable & Value(s) \\
		\hline
		$n_0$ & $10^{-4}$ cm$^{-3}$ \\
		$T_0$ & $10^7$ K \\
		Mach number & $0.9$, $1.1$, $1.9$ \\
		\hline
	\end{tabular}
\end{table}

\begin{table*}
\caption{Parameters for the simulations. $U_{1}$ is the speed of the ICM relative to the galaxy. $a_{1}$ and $a_{2}$ are the sound speeds in the ICM and in the galaxy at the stripping radius, respectively. The temperature and sound speed of the hot halo increases with radius. $\rho_{2}$ is the gas mass density of the galaxy at the stripping radius, and $\rho_{1}$ is the density of the ICM. $U_{\rm c}$ is the predicted convective velocity of the gas in the shear-layer (Eq.~\ref{eq:Uc}). $\delta_{0}'$ is the incompressible growth rate of the mixing layer (Eq.~\ref{eq:delta_vis_0}), $\pi_{\rm c}$ is a compressibility parameter (Eq.~\ref{eq:pic}), $\delta'(\pi_{\rm c})/\delta_{0}'$ is the normalized growth rate (Eq.~\ref{eq:normalized_growth}), and $\theta'(\pi_{\rm c})$ is the opening angle (Eq.~\ref{eq:opening_angle}). For the Mach 1.1 and 1.9 models, the ICM first passes through a bowshock upstream of the galaxy. The second line for each of these models therefore contains values computed with the free-streaming ICM conditions replaced by the conditions of the immediate post-shock flow on the symmetry axis. We have assumed that the stripping radius is unchanged and therefore do not adjust the values of $a_{2}$ and $\rho_{2}$.}
\label{table:3}
\centering
\begin{tabular}{lccccccccccc}
\hline
Model & $U_{1}$& Stripping & $a_{1}$ & $a_{2}$ & $\rho_{2}$ & $\rho_{2}/\rho_{1}$ & $U_{\rm c}$ & $\delta_{0}'$ & $\pi_{\rm c}$ & $\frac{\delta'(\pi_{\rm c})}{\delta_{0}'}$ & $\theta'(\pi_{\rm c})$ \\
Mach No. & (${\rm km\,s^{-1}}$) & radius (kpc) & (${\rm km\,s^{-1}}$) & (${\rm km\,s^{-1}}$) & (${\rm g\,cm^{-3}}$) & & (${\rm km\,s^{-1}}$) & & & &  ($^{\circ}$) \\
\hline
1.9 & 1000 & 120 & 526 & 320 & $4.6\times10^{-28}$ & 5.5  & 298 & 0.57 & 2.55 & 0.19 & 6.2 \\
    &  460 &     & 733 &     &                    & 2.5  & 346 & 0.44 & 1.17 & 0.39 & 9.8 \\
1.1 &  580 & 202 & 526 & 380 & $1.4\times10^{-28}$ & 1.65 & 254 & 0.39 & 1.25 & 0.37 & 8.2 \\
    &  504 &     & 550 &     &                    & 1.44 & 232 & 0.37 & 1.08 & 0.42 & 8.9 \\
0.9 &  470 & 236 & 526 & 425 & $0.9\times10^{-28}$ & 1.13 & 228 & 0.35 & 0.90 & 0.49 & 9.8 \\
\hline
\end{tabular}
\end{table*}

We take the galaxy to be comprised of a dark matter halo and a hot extended gas component. We followed \citet{2008MNRAS.383..593M} for the dark matter and gas density distributions. We used a NFW distribution  \citep*{1996ApJ...462..563N} for the dark matter:

\begin{equation}
	\rho(r)_{dm} = \frac{\rho_s}{(r/r_s)(1+r/r_s)^2}.
\end{equation}
The characteristic density of the dark matter halo is

\begin{equation}
	\rho_s = \frac{M_{s}}{4\pi r_s^3},
\end{equation}
where

\begin{equation}
	M_{s} = \frac{M_{200}}{ln(1+c_{200}) - c_{200}/1+c_{200}},
\end{equation}
$c_{200} = r_{200}/r_s$ is the concentration parameter and we take $c_{200} = 4$ in accordance with \citet{2007MNRAS.376..497M}. $r_{200}$ is the radius at which the average density is 200 times the critical density, $\rho_{crit} = 3 H_0^2 / 8 \pi G$. $M_{200}$ is the mass within $r_{200}$. We assumed the dark matter causes a static gravitational field. In the absence of tidal forces we do not expect the dark matter distribution to be significantly altered; so a static field is justified. The dark matter extends to a radius of $r_{25} \approx 2.44\,r_{200}$.

The initial hot gas component was assumed to follow the dark matter distribution, such that 

\begin{equation}
	\rho_g(r) = f_b \rho_t(r),
\end{equation}
where $\rho_b$ is the gas density, $\rho_t$ is the total density and $f_b = 0.022h^{-2}/0.3 = 0.141$ is the universal fraction of baryonic matter. The gas distribution of the galaxy is truncated at $r_{200}$. The temperature and density of the ICM were set to $10^7\,$K and $ 10^{-4}\,$cm$^{-3}$. Hydrostatic equilibrium of the halo gas was assumed and determines the temperature and pressure distributions. The gas contained within $r_{200}$ was given an advected scalar so that mass from the galaxy could be identified at later times. The total initial mass of the hot gaseous halo is $5.64\times10^{11}\,{\rm M_{\odot}}$. The galaxy was allowed to evolve in isolation with and without the $k$-$\epsilon$ model implemented and is stable for many Gyr.

The  initial  values of  $k$  and $\epsilon$  were  set  to $3  \times
10^{10}\,$erg$\,$g$^{-1}$  and to $3  \times 10^{10}\,$erg$\,$g$^{-1}$
respectively. As  these values are  several orders of  magnitude lower
than those  subsequently generated during the  interaction, the effect
of the turbulent viscosity is  dominated by the real turbulence at all
times. We note that a  positive feature of the $k$-$\epsilon$ model is
its  relative insensitivity  to  the initial  conditions  for $k$  and
$\epsilon$ (see e.g. Section 5.4.3 of \citet{Wilcox:2006}).

The  ICM was given  an initial  velocity $U_{1}$  in the  positive $x$
direction to simulate the effects  of the galaxy moving though it. The
lower  $x$  boundary is  set  to  drive the  wind  at  the same  speed
throughout the simulation. In reality  a galaxy falling into a cluster
will  see a  gradual  increase in  wind  speed.  This  will cause  the
instantaneous stripping phase to be  drawn out over a longer period of
time. By  the onset  of continual stripping  however, enough  time has
passed ($\sim  2\,$Gyr) that any  shocks caused by the  initial impact
have dissipated.  Three different velocities  were used, corresponding
to   Mach   numbers   of    $0.9$,   $1.1$   and   $1.9$   ($U_{1}   =
470\,$km$\,$s$^{-1}$, $580\,$km$\,$s$^{-1}$ and $1000\,$km$\,$s$^{-1}$
respectively). These values are typical of those in the literature and
representative of the orbital  speeds of a cluster galaxy. Simulations
for each  different Mach  number were performed  with and  without the
$k$-$\epsilon$ model.

\subsection{Parametrization of Galaxy Mass}
\label{sec:gmass}

In order to measure the mass of the gaseous component of the galaxy as a function of time, and hence the effectiveness of ram pressure stripping, we track the mass in four ways. $M_{\mathrm{g,r200}}$, the mass of gas that was initially part of the galaxy and is within a distance of $r_{200}$ of the galactic centre. $M_{\mathrm{g,unmixed}}$ the mass of gas that was initially in the galaxy and that is in cells in which as least half of the gaseous material started the simulation in the galaxy. $M_{\mathrm{g,bound,gal}}$, the mass of gas that was initially in the galaxy that is gravitationally bound to the dark matter potential. $M_{\mathrm{g,bound}}$, the total mass of gas that is bound to the dark matter potential.

\subsection{Effects of Resolution}

\begin{figure}
	\centering
	\includegraphics[width=0.5\textwidth]{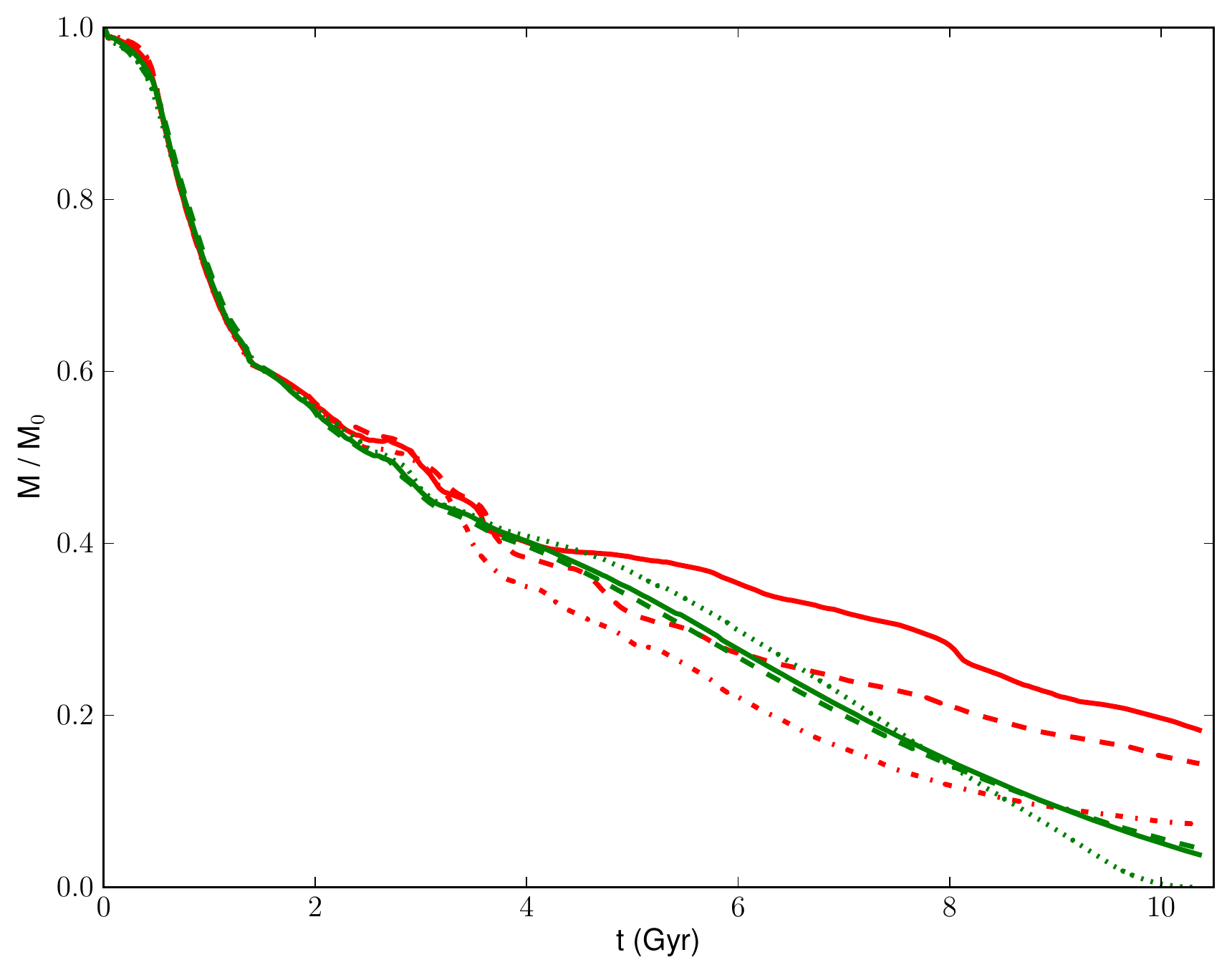}
	\caption{$M_{\mathrm{g,bound}}$  as a function  of time  for a
	Mach number of $1.9$.  The results for the model incorporating
	the  $k$-$\epsilon$ model are  shown in  green. Those  for the
	inviscid  model  are in  red.  The  solid  lines are  for  the
	standard resolution  used throughout  the rest of  this study,
	the  dashed lines  are for  the runs  with an  extra  level of
	refinement, which allowed twice the maximum resolution and the
	dotted  line  uses  one  less refinement  level,  halving  the
	maximum  resolution.  The   dash-dotted  line  has  two  extra
	refinement  levels,  increasing the  maximum  resolution by  a
	factor of four to 1.5kpc.}
	\label{fig:res}
\end{figure}

\begin{figure*}
	\centering
	\includegraphics[width=0.9\textwidth]{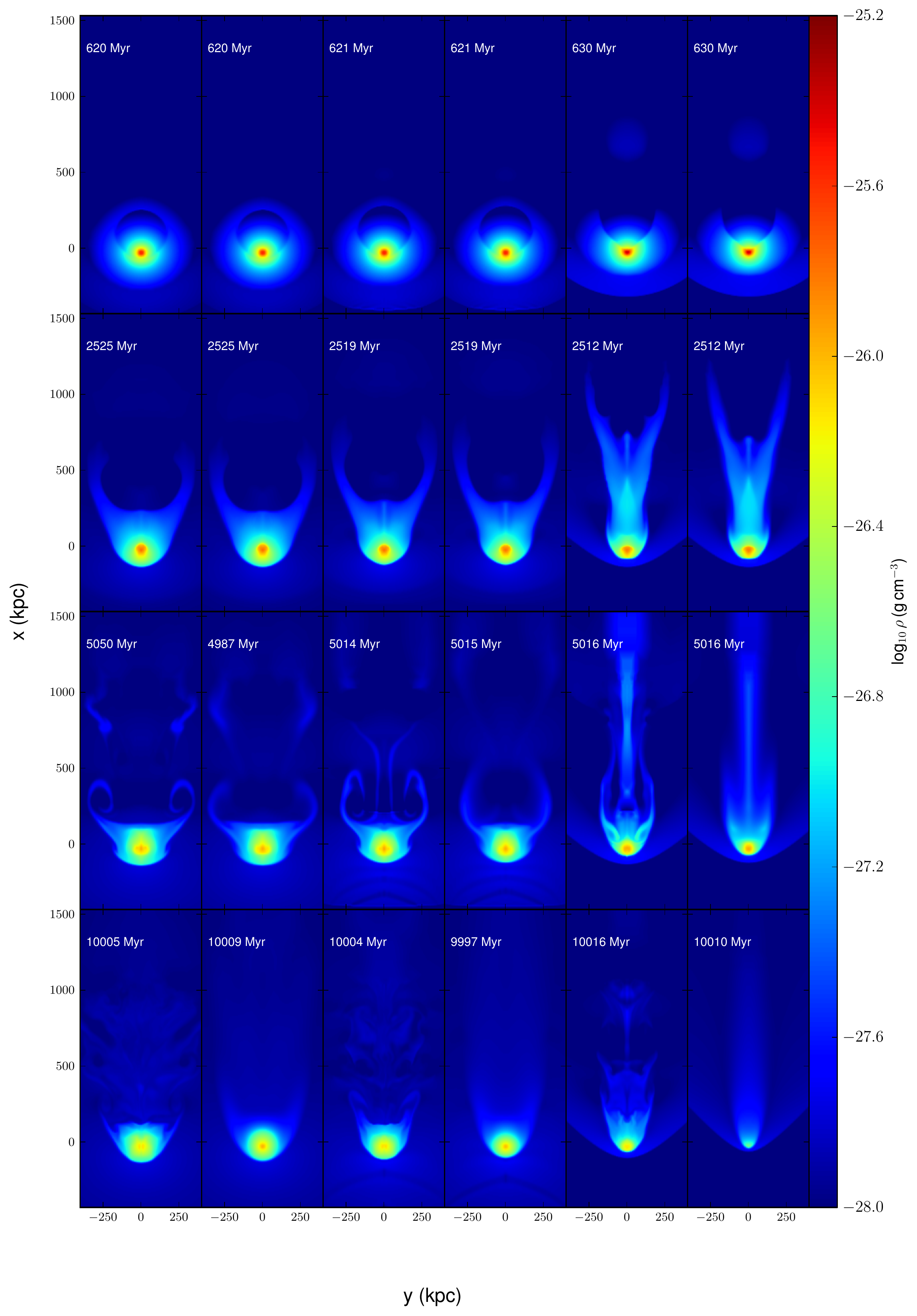}
	\caption{The mass density in the $z=0$ plane. The first two columns contain results for simulations at a Mach number of $0.9$, the second two results for a Mach number of $1.1$ and the last two results for a Mach number of $1.9$. The left column in each pair shows results for the inviscid simulations, while the right column in each pair gives results for models that incorporate the $k$-$\epsilon$ model.}
	\label{fig:rho}
\end{figure*}

\begin{figure*}
	\centering
	\includegraphics[width=\textwidth]{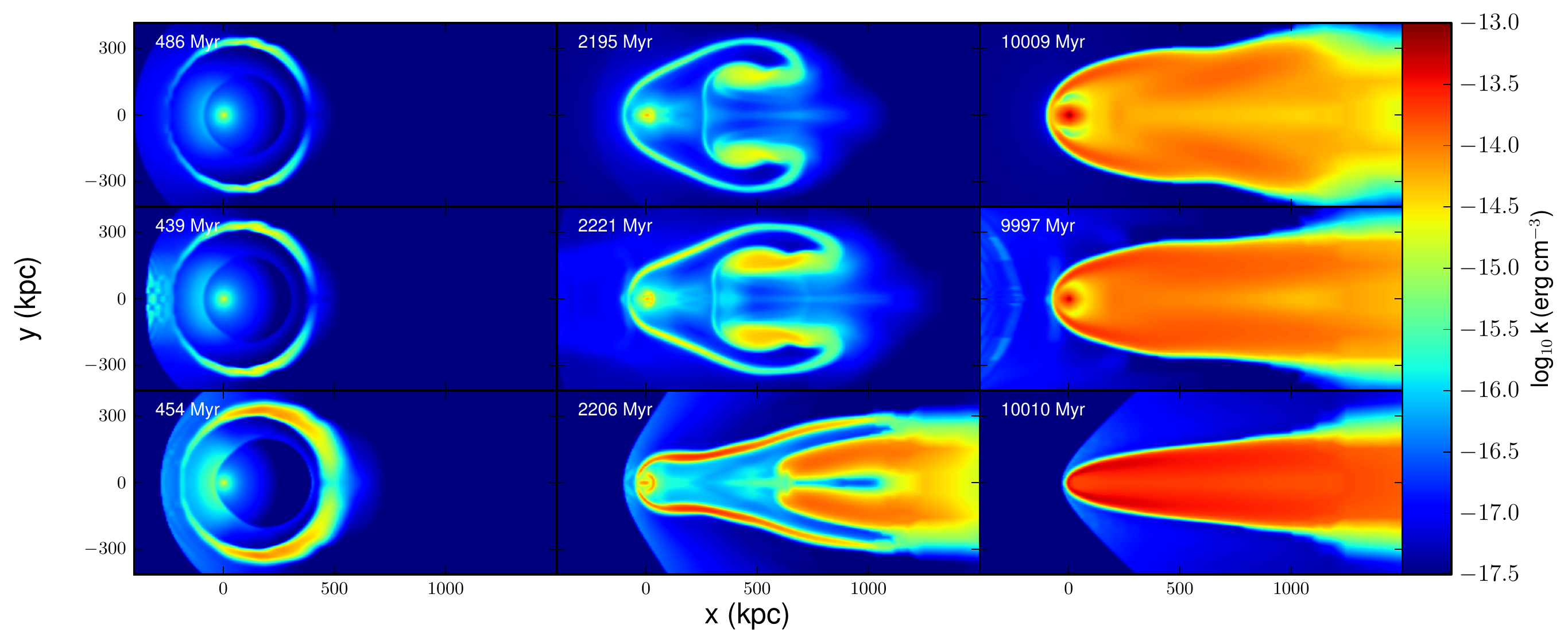}
	\caption{The distribution in the $z=0$ plane of turbulent energy, $k$, for simulations incorporating the $k$-$\epsilon$ model. Rows correspond to, from the top, Mach numbers of $0.9$, $1.1$ and $1.9$. The three times are: during instantaneous stripping, after instantaneous stripping and at the end of the simulation.}
	\label{fig:kslices}
\end{figure*}

In order to investigate the  effects of resolution, we ran the highest
Mach number simulation with one extra refinement level. The bound mass
for these tests is shown  in fig.  \ref{fig:res}.  At early times both
methods  give  similar results,  which  is  not  surprising since  the
instantaneous stripping is unaffected by turbulence. At later time the
$k$-$\epsilon$ model converges, whereas there  is no evidence
of convergence in the inviscid runs.  The inviscid calculations can be
regarded  as  Implicit  Large  Eddy Simulations  since  the  numerical
algorithm  has the  required  properties (\citealt{Aspden:2008}),  but
such  calculations are  clearly not  reliable unless  one  can achieve
convergence. This we  have been unable to do even  though we have used
higher   resolution   than    in   previous   calculations   such   as
\citet{2013MNRAS.428..804S}.

The advantage of the $k$-$\epsilon$ model is that it converges
at a  resolution that can be  used in practical  calculations. We also
have shown results for a run  with one less refinement level using the
$k$-$\epsilon$ model.  At this resolution the results begin to diverge
from the higher resolution runs.


\section{Analytical Approximations}

\subsection{Instantaneous Stripping}

Instantaneous stripping occurs when the ram pressure exceeds the gravitational force per unit area on a column of material. In \citet{2008MNRAS.383..593M} an analytical prediction for the instantaneous ram pressure stripping of a spherically symmetric galaxy was derived, which extended work done by \citet{1972ApJ...176....1G}. The condition for ram pressure stripping at a given radius, $R$ is, 

\begin{equation}
	\rho_{icm} \nu_{gal}^2 > \alpha\frac{GM_{gal}(R)\rho_{gas}(R)}{R},
	\label{eqn:is}
\end{equation}
where $\alpha$ is a constant dependent on the dark matter and gas profiles. \citet{2008MNRAS.383..593M} found that for their wind tunnel tests $\alpha \approx 2$ best fits their results. The stripping radius for each model is noted in Table~\ref{table:3}. We also note the mass density and the sound speed of the hot gaseous galactic halo at this radius ($\rho_{2}$ and $a_{2}$ in Table~\ref{table:3}, respectively).

\subsection{Kelvin-Helmholtz Stripping}

The rate at which mass is lost due to the Kelvin-Helmholtz instability was estimated by \citet{1982MNRAS.198.1007N} to be

\begin{equation}
	\dot{M}_{kh} = \pi r^2 \rho_{icm} \nu_{gal}.
	\label{eqn:khs}
\end{equation}

Taking $r$ as the radius after instantaneous stripping from equation (\ref{eqn:is}) and using $1000\,$km$\,$s$^{-1}$ as the velocity of the galaxy we get a mass loss rate of $49\,$M$_{\sun}\,$yr$^{-1}$ for the assummed galaxy. At this rate the hot halo would be completely stripped in $4.4\,$Gyr. We might expect this to be the rate at which mass is lost initially, but as mass is stripped the radius will decrease. One can assume that the mass-radius relation remains constant in order to calculate how the radius changes as mass is removed from the galaxy.

\subsection{Compressible Turbulent Shear Layers}

Turbulent shear layers are also of wide interest to the fluid-mechanics community and much study has been devoted to them. \citet{Brown:1974} determined that the visual thickness of the shear layer spread as
\begin{equation}
\label{eq:delta_vis_0}
\delta_{\rm vis,0}' = \frac{\delta(x)}{x - x_{\rm 0}} \approx 0.17\,\frac{\Delta U}{U_{\rm c}},
\end{equation}
where $\delta(x)$ is the thickness of the mixing layer at a downstream distance of $x - x_{\rm 0}$ from the point where the streams initially interact, $\Delta U = U_{\rm 1} - U_{\rm 2}$ is the velocity difference between the two free streams and $U_{\rm c}$ is the ``convective velocity'' at which large-scale eddies within the mixing layer are transported downstream. The constant 0.17 was empirically determined. When stream ``2'' is stationary, 
\begin{equation}
\label{eq:Uc}
U_{\rm c} = \frac{U_{\rm 1}}{1 + \left(\frac{\rho_{2}}{\rho_{1}}\right)^{1/2}},
\end{equation}
where the densities of the free streams are $\rho_{1}$ and $\rho_{2}$ \citep[e.g.,][]{Papamoschou:1988}. A slightly more complicated form of Eq.~\ref{eq:delta_vis_0} is noted in \citet{Soteriou:1995}.

It has long been recognized that the growth rate of compressible mixing layers is lower than predicted by Eq.~\ref{eq:delta_vis_0} when the convective velocity is high. This is attributed to compressibility effects. \citet{Papamoschou:1988} showed that experimental measurements of the growth rate normalized by its incompressible value, $\delta'/\delta_{\rm 0}'$, was largely a function of the ``convective Mach number'', $M_{\rm c,1} = (U_{1} - U_{\rm c})/a_{1}$, where $a_{1}$ is the sound speed of free stream ``1''. More recently, \citet{Slessor:2000} argue that the convective Mach number under-represents compressibility effects for free streams with a significant density ratio, and show that a better characterisation is achieved through the use of an alternative compressibility parameter,
\begin{equation}
\label{eq:pic}
\pi_{\rm c} = {\rm max}_{i} \left[\frac{\sqrt{\gamma_{i} - 1}}{a_{i}}\right] \Delta U,
\end{equation}
where $\gamma_{i}$ and $a_{i}$ are the ratio of specific heats and the sound speed of free stream {\it i}. For $\pi_{\rm c} \ltsimm 3$, the normalized growth rate
\begin{equation}
\label{eq:normalized_growth}
\frac{\delta'(\pi_{\rm c})}{\delta_{\rm 0}'} \approx \left(1 + \alpha \pi_{\rm c}^{2}\right)^{-1/2}.
\end{equation}
The opening angle of the mixing layer is then
\begin{equation}
\label{eq:opening_angle}
\theta'(\pi_{\rm c}) = 2\arctan\left(\frac{\delta'(\pi_{\rm c})}{2}\right).
\end{equation}


\section{Results}
\subsection{Wind tunnel tests}

\begin{figure}
	\centering
	\includegraphics[width=0.5\textwidth]{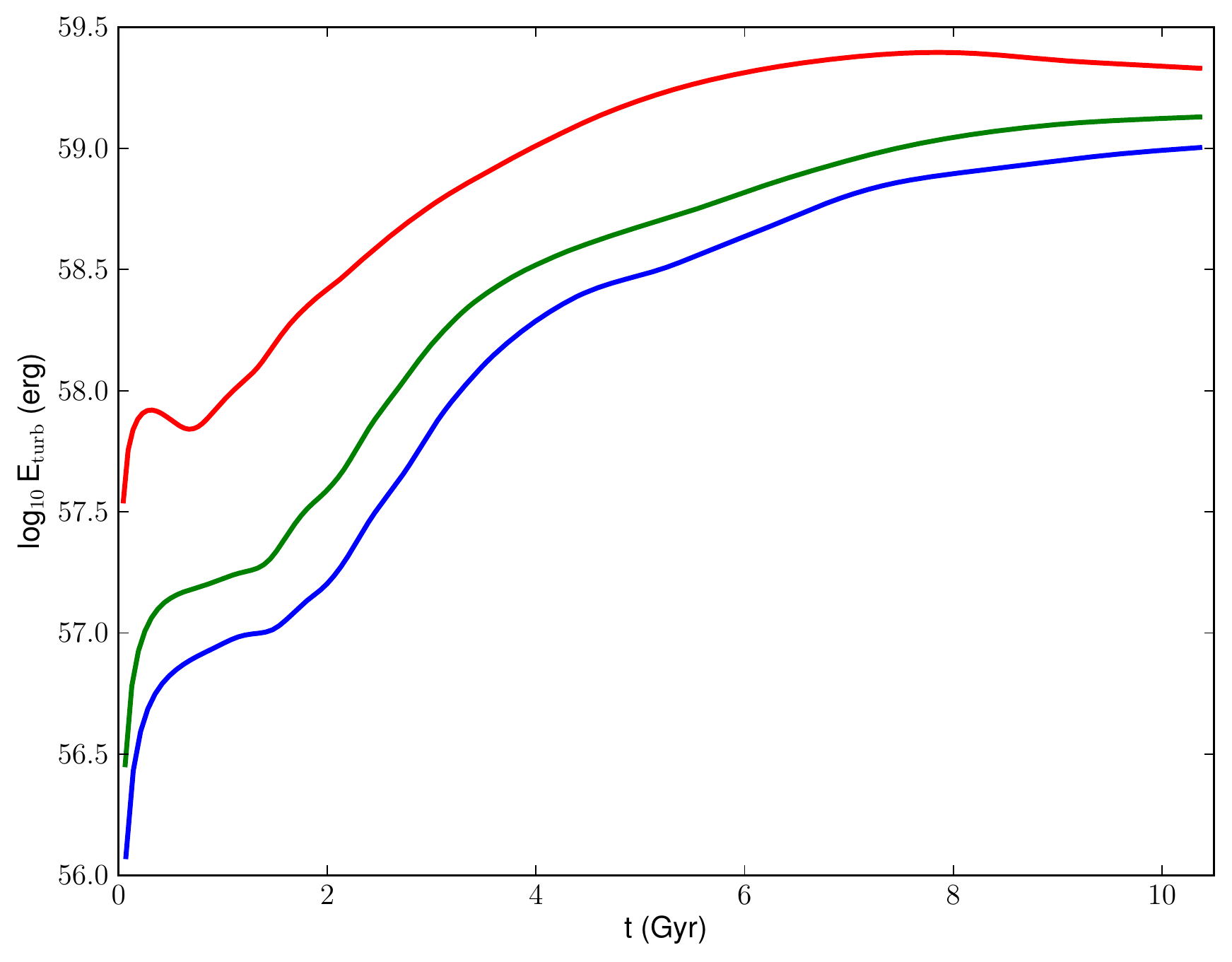}
	\caption{The total turbulent energy, $E_{\mathrm{turb}}$, across the entire grid. The blue, green and red lines give results for the simulations with Mach numbers of $0.9$, $1.1$ and $1.9$ respectively.}
	\label{fig:k}
\end{figure}

\begin{figure*}
	\centering
	\includegraphics[width=0.8\textwidth]{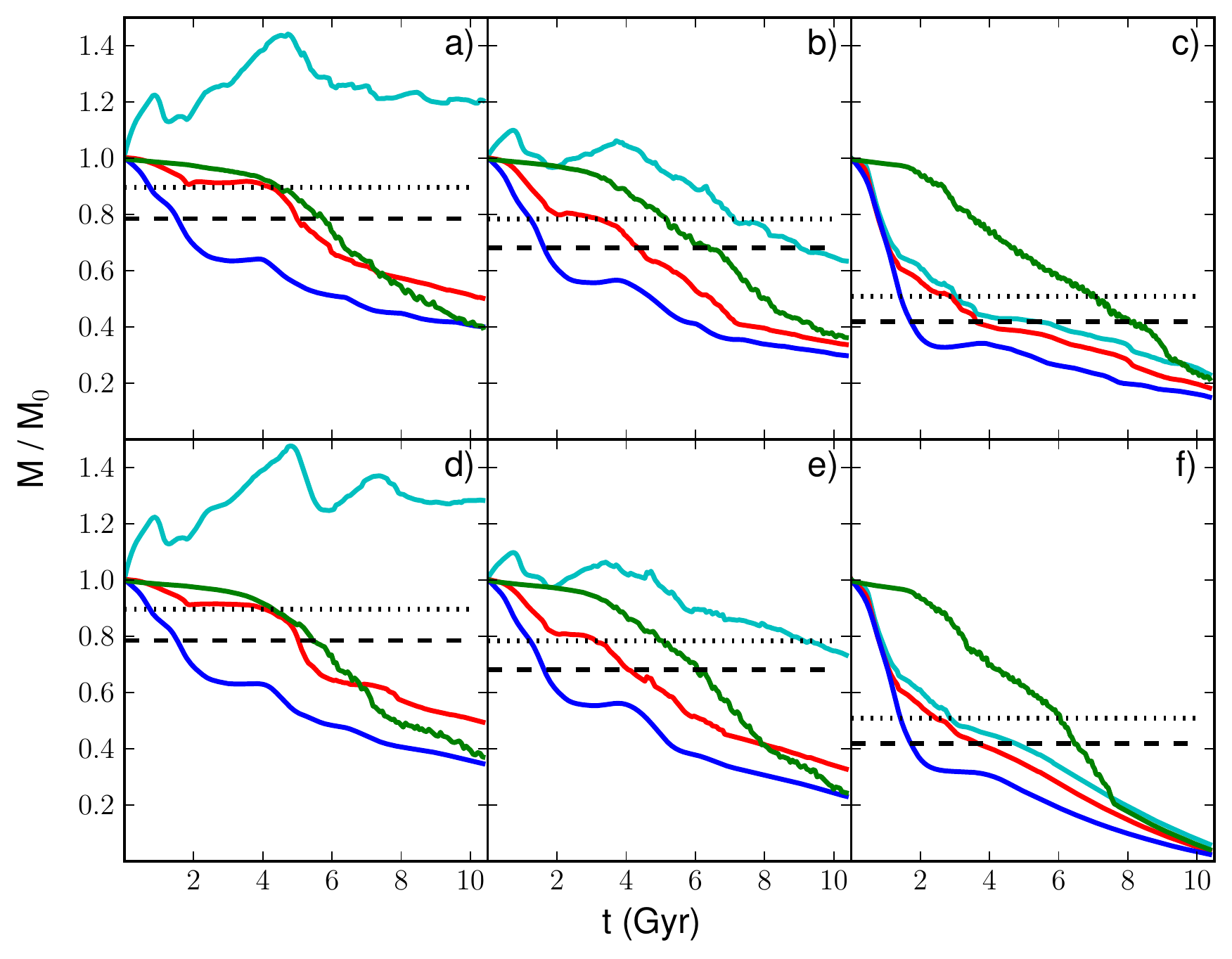}
	\caption{Gaseous component of the galaxy mass as a function of time for all six simulations. Results for inviscid runs are shown in a-c, and results for runs incorporating the $k$-$\epsilon$ model in d-f. Columns correspond to, from the left, Mach numbers of $0.9$, $1.1$ and $1.9$. The blue line shows $M_{\mathrm{g,r200}}$, the green line shows $M_{\mathrm{g,unmixed}}$, the red line shows $M_{\mathrm{g,bound,gal}}$ and the cyan line shows $M_{\mathrm{g,bound}}$. The black dashed line is the analytical prediction from equation (\ref{eqn:is}) with $\alpha = 2$. The black dotted line gives the result for the same equation but for $\alpha = 3$.}
	\label{fig:mass}
\end{figure*}

\begin{figure}
	\centering
	\includegraphics[width=0.5\textwidth]{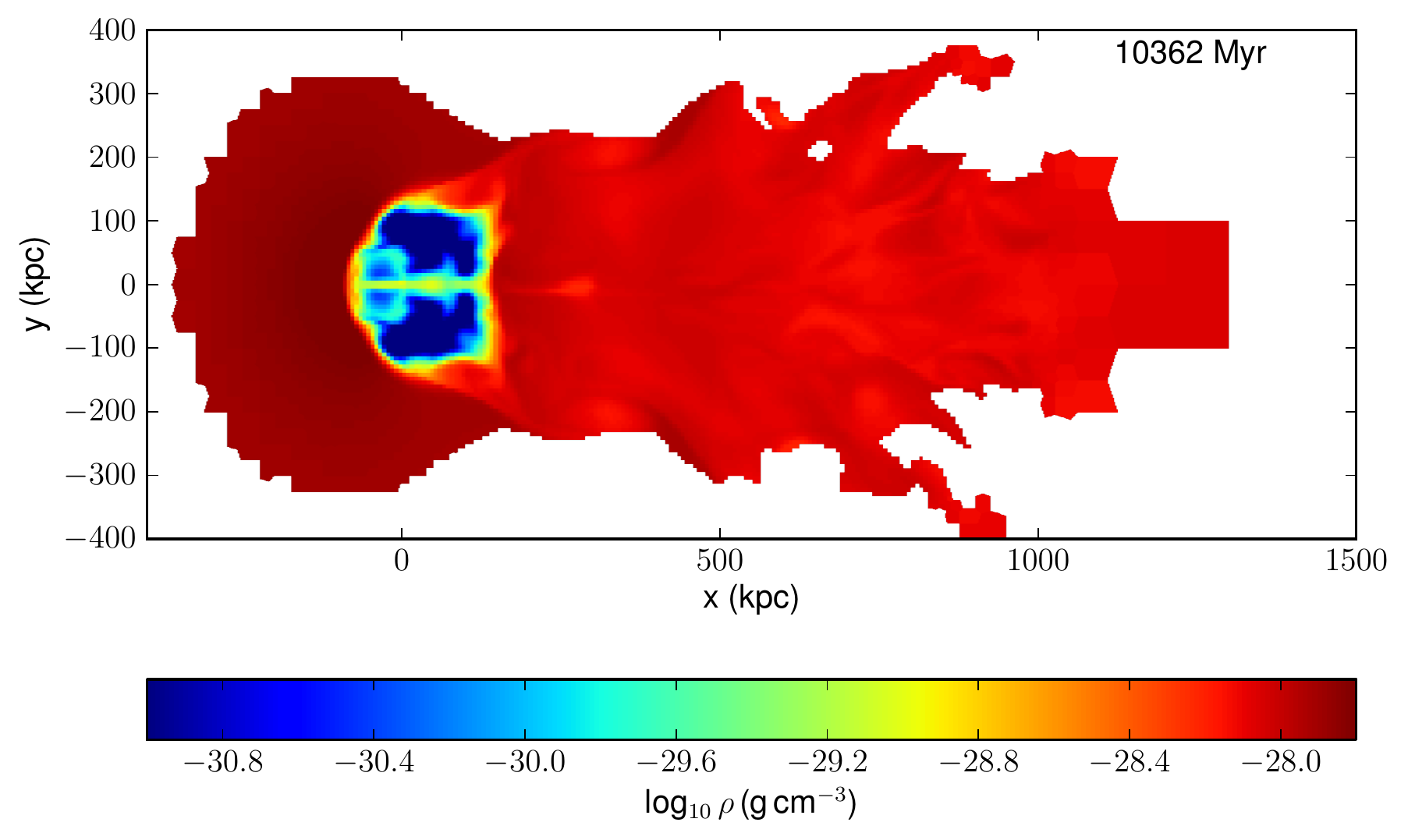}
	\caption{The distribution of gas that started the simulation as ICM and is bound to the dark matter potential, for the 0.9 Mach number model incorporating the $k$-$\epsilon$ model. In the blue area the gas bound to the galaxy is mostly gas that was associated with the galaxy initially.}
	\label{fig:nongal}
\end{figure}

\begin{figure}
	\centering
	\includegraphics[width=0.5\textwidth]{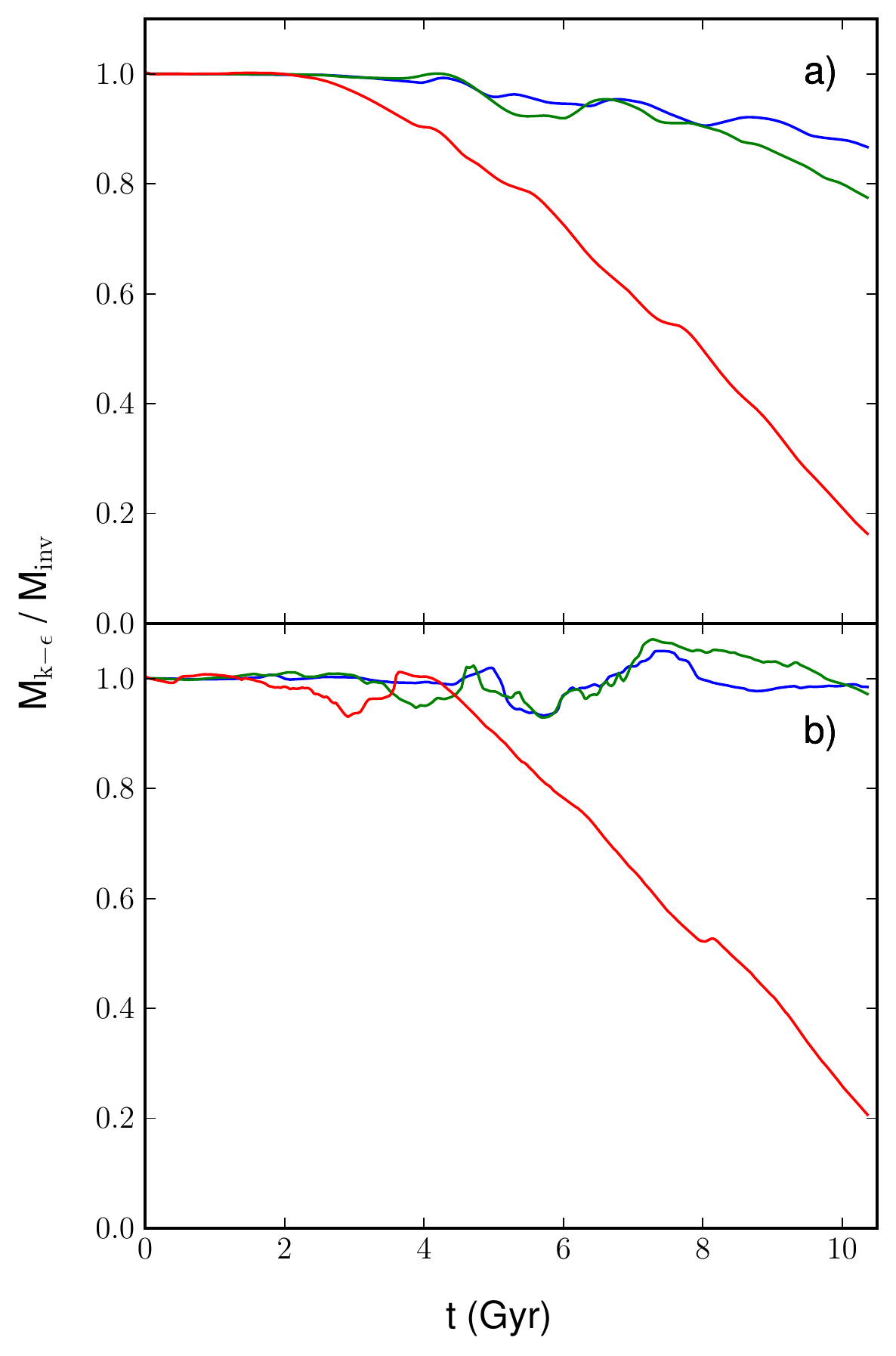}
	\caption{Ratio of the mass of the galaxy in $k$-$\epsilon$ runs over the corresponding inviscid run. The top panel shows $M_{\mathrm{g,r200}}$, while the bottom panel $M_{\mathrm{g,bound,gal}}$. The blue, green and red lines correspond to Mach numbers of $0.9$, $1.1$ and $1.9$ respectively.}
	\label{fig:massratio}
\end{figure}

\begin{figure}
	\centering
	\includegraphics[width=0.5\textwidth]{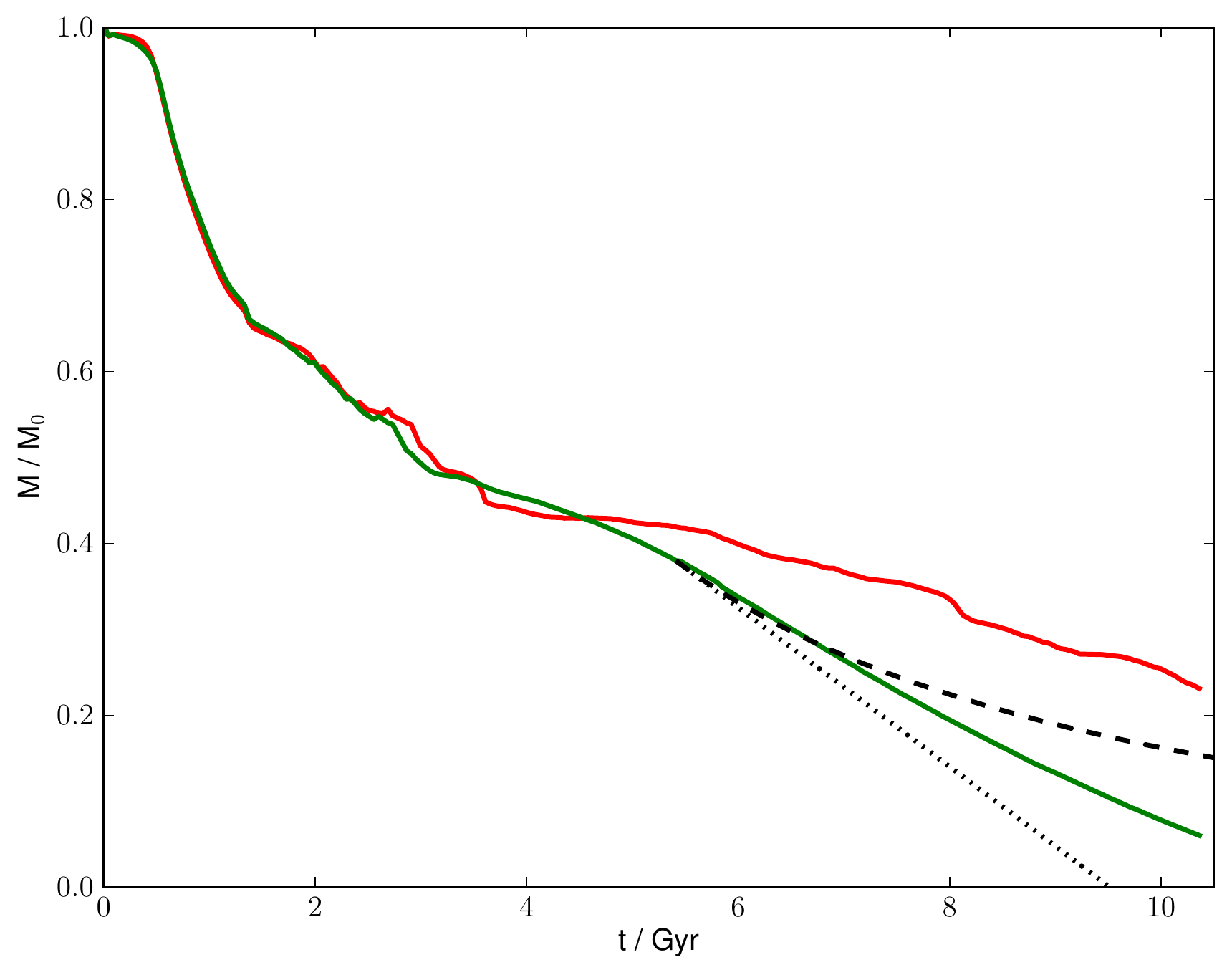}
	\caption{$M_{\mathrm{g,bound}}$ as a function of time for a Mach number of $1.9$. Shown in green is the run incorporating the $k$-$\epsilon$ model and in red is the inviscid run. The black dotted line shows the prediction from equation (\ref{eqn:khs}), assuming a constant radius. The black dashed line shows the same equation but allowing the radius to change over time.}
	\label{fig:kh}
\end{figure}

Fig. \ref{fig:rho} shows snapshots of the mass density distribution at a number of times for each of the runs. The $k$-$\epsilon$ runs agree relatively closely with their counterparts during the instantaneous stripping phase, but become increasingly divergent at later times when the Kelvin-Helmholtz instability is most important. The largest difference between the inviscid and $k$-$\epsilon$ simulations are seen in the Mach $1.9$ run. The results for runs incorporating the $k$-$\epsilon$ model look smoother in general as the density given is a local mean density.

Fig. \ref{fig:kslices} shows the  distribution of turbulent energy for
each $k$-$\epsilon$  run. The turbulence is initially  produced at the
interface between the galaxy and  the ICM. There is also turbulence in
the tail after instantaneous stripping,  as seen in the central column
of panels. This is  initially generated during instantaneous stripping
and is advected downstream. At later times the turbulence generated at
the galaxy-ICM  interface has managed  to propagate into and  fill the
tail. Some of  the turbulence in the tail also  moves back towards the
galaxy.  The general  trend  of increased  turbulence  at higher  Mach
numbers    is   apparent.    This   is    shown    quantitatively   in
fig. \ref{fig:k}. At sonic  and marginally supersonic Mach numbers the
turbulent energy rises  gradually with time after an  initial delay of
$\sim2\,$Gyr  while instantaneous  stripping is  occurring.  The total
turbulent  energy peaks  at $\sim8\,$Gyr  in the  highest  Mach number
simulation  as turbulence  begins to  be advected  from the  grid. The
typical value of  the turbulent diffusivity in the  shearing layers is
$\mu_T \sim 3 \times  10^{28}\,$cm$^2$s$^{-1}$ (as defined by equation
\ref{eqn:td}).

Fig. \ref{fig:mass} shows each of the different ways we tracked the mass of the galaxy, as described in section \ref{sec:gmass}. In the first $\sim2\,$Gyr gas is removed where gravity is weak enough to be overcome by ram pressure, and $M_{g,bound,gal}$ (red) and $M_{\mathrm{g,r200}}$ decreases. $M_{\mathrm{g,unmixed}}$ (green) remains relatively constant during this period, showing that the ram pressure is not mixing the gases, but rather pushing it from the galaxy, as we might expect. This occurs on a time-scale equal to the sound crossing time ($1.8\,$Gyr). It should be noted that in the transonic cases much of the ICM component of $M_{\mathrm{g,bound}}$ is in front of the galaxy and is still part of the flow. For example, in the Mach 0.9 case (shown in fig. \ref{fig:nongal}), the wind speed is equal to the escape velocity at $\sim$$170\,$kpc, so gas within this radius is technically bound but is being constantly pushed downstream due to the flow of the wind behind it. After this the galaxy enters a transitional period where little to no mass is stripped from the galaxy, particularly at lower Mach numbers. The galaxy then begins to be stripped by the Kelvin-Helmholtz instability until the end of simulation and the mass associated with the galaxy decreases.

The general shape of the curves is the same in all cases (with the exception of $M_{\mathrm{g,bound}}$, as noted above). The lengths of the instantaneous stripping and transitional periods appear independent of Mach number, with the transitional period being more pronounced at lower Mach numbers. In general stripping is stronger at higher Mach numbers. In all cases the galaxy loses more mass than predicted due to instantaneous stripping alone. This means that Kelvin-Helmholtz stripping is significant on the $10\,$Gyr time-scales investigated. The simulations incorporating the $k$-$\epsilon$ model differ the most from the corresponding inviscid simulation at the highest Mach number (fig. \ref{fig:mass}c and \ref{fig:mass}f), particularly in the late-time Kelvin-Helmholtz stripping phase, when the turbulence is fully developed.

Fig. \ref{fig:massratio} shows the difference in mass stripping between the inviscid and the $k$-$\epsilon$ simulations in more detail. While there is little difference in $M_{\mathrm{g,bound,gal}}$ at the transonic Mach numbers (but some difference in $M_{\mathrm{g,r200}}$) at supersonic Mach numbers we see that there is significantly more stripping when the $k$-$\epsilon$ model is used. For the Mach 1.9 case, $M_{\mathrm{g,r200}}$ and $M_{\mathrm{g,bound,gal}}$ are both only 20\% of the mass in the inviscid model at late times.

Fig. \ref{fig:kh} shows the simple analytical predictions of equation (\ref{eqn:khs}) for a Mach number of $1.9$. The galaxy undergoes a transitional period between the instantaneous and Kelvin-Helmholtz phases thus the time at which results should be expected to follow the analytic prediction is unclear. For any chosen starting point after $\sim5\,$Gyr both curves fit the simulation well initially and slowly diverge away. This suggests that although the simulation strips at a rate which is similar to the analytical prediction, the formula is an over simplification of the time dependence of the radius.

Table~\ref{table:3} notes predicted values for the convective velocity, $U_{\rm c}$, for each simulation. However, it is not possible for us to easily compare our simulation results to this value because our simulations provide us with the mean local velocity instead, and this changes across the mixing layer from the undisturbed galactic halo gas to the unmixed ICM gas flowing past. In addition, \citet{Papamoschou:1997} show that predicitions for the convective Mach number and the convective velocity need correcting if the mixing is not ``symmetric''. For these reasons we choose to instead focus on the opening angle of the mixing layer. The predicted openging angles for the Mach 0.9 model and the post-shock-adjusted predictions for the Mach 1.1 and 1.9 models yield $\theta'(\pi_{\rm c}) \approx 10^{\circ}$. This compares with the observed angle of about 10$^\circ$-15$^\circ$ estimated from Fig.~\ref{fig:k}, and is also consistent with the results of \citet{1991ApJ...372..646C} (see also the end of section 4.2 in \citealt{2009MNRAS.394.1351P}). We consider this level of agreement as perfectly acceptable given that our simulations differ in some noteable respects from the experiments we are comparing against. For instance, in our work i) the mixing layer is not flat but instead curves around the galaxy; ii) the ICM gas does not stream past the galaxy at a steady velocity (it is nearly stationary close to the stagnation point of the flow upstream of the galaxy and accelerates around the galaxy); iii) the gas experiences gravitational forces from the galaxy. We conclude that these differences affect the opening angle of the mixing layer by less than a factor of two.

\subsection{Comparison to previous works}

\citet{2008MNRAS.383..593M} investigated ram pressure stripping of a hot galactic halo with the SPH code \texttt{GADGET-2}. Our bound mass after instantaneous stripping lies above what one would expect from equation (\ref{eqn:is}) using the $\alpha \sim 2$ value that they determined from their results, particularly at lower Mach numbers, which they did not investigate. The method they used for calculating bound mass is fundamentally different from ours (due to the differences in SPH and grid-based codes) and they discussed this in appendix A of their paper. They found that implementing the same bound mass calculation as we use increases the amount of mass bound after instantaneous stripping suggesting that this could be the reason for the discrepancy between our results. Our results are more consistent with $\alpha = 3$, as shown in fig. \ref{fig:mass}. They noted that instantaneous stripping occurs on time-scales comparable to the sound crossing time of the galaxy, which is what we found. They did not however see any mass loss after the instantaneous stripping period, in contrast to our results.

\cite{2009MNRAS.399.2221B} included a disc in the galaxy in addition to a halo and used the SPH code \texttt{GRAPE-SPH}. He only conducted two-body tests (i.e. no wind tunnel tests) so a direct comparison with our results is not possible. However he remarked that his results are broadly consistent with those of \citet{2008MNRAS.383..593M}. He also does not see any continual stripping after the initial mass loss.

Most recently \citet{2013MNRAS.428..804S} used the grid-based code \texttt{FLASH3} to study ram pressure stripping of elliptical galaxies. They measured the mass of their galaxy by measuring the mass within radial zones and so their results are most comparable with our $M_{\mathrm{g,r200}}$ calculations. They look specifically at subsonic galaxy speeds (Mach number of 0.25) so our results are not necessarily directly comparable but we see broadly the same pattern of quick, instantaneous stripping, followed by a plateau and then continual stripping to the end of simulation.


\section{Conclusion}

We have studied the effects of ram pressure stripping on the halo of a massive galaxy using three dimensional hydrodynamic simulations incorporating the $k$-$\epsilon$ sub-grid turbulence description. We have varied the Mach number of the interaction to investigate its effect.

We found that in all cases the Kelvin-Helmholtz instability contributes significantly to the stripping of material from the galaxy. This is captured both with and without use of the $k$-$\epsilon$ model at transonic Mach numbers. During instantaneous stripping the simulations incorporating the $k$-$\epsilon$ model produces the same results as simulations in which it is not used. However at higher Mach numbers (i.e. $1.9$) Kelvin-Helmholtz stripping is only properly captured when the $k$-$\epsilon$ model is used. 

This means that the stripping of gas from hot gaseous haloes has been underestimated in previous simulations particularly those in which the galaxy Mach number is about two or more. Since it is currently not feasible to accurately model fully developed turbulence in this problem, the incorporation of a sub-grid turbulence model, such as the $k$-$\epsilon$ model, is highly desirable.


\end{document}